\documentclass[conference]{IEEEtran}
\usepackage{booktabs}
\usepackage{multicol}% http://ctan.org/pkg/multicols
\usepackage{multirow}% http://ctan.org/pkg/multirow
\usepackage{graphicx}% http://ctan.org/pkg/graphicx
\usepackage{tabularx}% http://ctan.org/pkg/tabularx
\usepackage{float}

\usepackage{lipsum}
\usepackage{mathtools}
\usepackage{cuted}

\usepackage{cite}
\usepackage{amsmath,amssymb,amsfonts}
\usepackage{algorithmic}
\usepackage{graphicx}
\usepackage{textcomp}
\usepackage{listings}
\usepackage{comment}
\usepackage{cuted}
\usepackage{algorithm}
\usepackage{xcolor}

\def\BibTeX{{\rm B\kern-.05em{\sc i\kern-.025em b}\kern-.08em
    T\kern-.1667em\lower.7ex\hbox{E}\kern-.125emX}}

\begin{document}

\title{Context-based smart contracts for \\appendable-block blockchains\thanks{
This paper was achieved in cooperation with HP Brasil using incentives of Brazilian Informatics Law (Law n 8.248 of 1991). This study was financed in part by the Coordena\c{c}\~ao de Aperfei\c{c}oamento de Pessoal de N\'ivel Superior - Brasil (CAPES) - Finance Code 001. Avelino F. Zorzo is supported by CNPq (315192/2018-6) and FAPERGS. This work was supported by the INCT Forensic Sciences through the Conselho Nacional de Desenvolvimento Cient\'ifico e Tecnol\'ogico (CNPq $-$ process \# 465450/2014-8). Also, we thank the support from IFRS and Cyber Security Research Centre Limited whose activities are partially funded by the Australian Government's Cooperative Research Centres Programme.}
}

% \author{\IEEEauthorblockN{Henry C. Nunes}
% \IEEEauthorblockA{PUCRS\\
% Porto Alegre, Brazil\\
% Email: henry.nunes@edu.pucrs.br}
% \and
% \IEEEauthorblockN{Roben C. Lunardi}
% \IEEEauthorblockA{IFRS and IFRS\\
% Porto Alegre, Brazil\\
% Email: roben.lunardi@edu.pucrs.br}
% \and
% \IEEEauthorblockN{Avelino F. Zorzo}
% \IEEEauthorblockA{PUCRS\\
% Porto Alegre, Brazil\\
% Email: avelino.zorzo@pucrs.br}}
\author{\IEEEauthorblockN{
Henry C. Nunes\IEEEauthorrefmark{1},
Roben C. Lunardi\IEEEauthorrefmark{1}\IEEEauthorrefmark{2},
Avelino F. Zorzo\IEEEauthorrefmark{1},
Regio A. Michelin\IEEEauthorrefmark{3}
 and
Salil S. Kanhere\IEEEauthorrefmark{3}}
\IEEEauthorblockA{\IEEEauthorrefmark{1}Pontifical Catholic University of Rio Grande do Sul (PUCRS) - Porto Alegre, Brazil}
\IEEEauthorblockA{ \IEEEauthorrefmark{2}Federal Institute of Rio Grande do Sul (IFRS) - Porto Alegre, Brazil}
\IEEEauthorblockA{\IEEEauthorrefmark{3}University of New South Wales (UNSW) - Sydney, Australia}
\IEEEauthorblockA{E-mail:  \{henry.nunes,roben.lunardi\}@edu.pucrs.br, avelino.zorzo@pucrs.br, \{regio.michelin,salil.kanhere@unsw.edu.au\}}}

\maketitle              % typeset the header of the contribution
\begin{abstract}
Currently, blockchain proposals are being adopted to solve security issues, such as data integrity, resilience, and non-repudiation. To improve certain aspects, \textit{e.g.}, energy consumption and latency, of traditional blockchains, different architectures, algorithms, and data management methods have been recently proposed. For example, appendable-block blockchain uses a different data structure designed to reduce latency in block and transaction insertion. It is especially applicable in domains such as Internet of Things (IoT), where both latency and energy are key concerns. However, the lack of some features available to other blockchains, such as Smart Contracts, limits the application of this model. To solve this, in this work, we propose the use of Smart Contracts in appendable-block blockchain through a new model called context-based appendable-block blockchain. This model also allows the execution of multiple smart contracts in parallel, featuring high performance in parallel computing scenarios. Furthermore, we present an implementation for the context-based appendable-block blockchain using an Ethereum Virtual Machine (EVM). Finally, we execute this implementation in four different testbed. The results demonstrated a performance improvement for parallel processing of smart contracts when using the proposed model.

\end{abstract}


\section{Introduction}\label{sec:intro}

Distributed applications, such as distributed databases \cite{Sheth:1990}, have existed for a long time. However, they are dependent on an assumption of trust, \textit{i.e.}  nodes that compose the environment are honest. Alternatively, trust can be delegated to a third party that can assure that the environment is trustable. One such third party is a certificate authority\cite{Berkowsky:2017} that can assure the identity of the nodes. The blockchain concept changed this scenario, ensuring the robust execution of a deployed application even if some of the participating nodes misbehave or become unavailable. Blockchain also offers other benefits, including audibility, transparency, and the possibility to have decentralized applications (dApps). The combination of those benefits allows to use blockchain in multiple domains, such as financial operations\cite{Zheng:2017}, supply chain~\cite{Niya:2019}, Internet of Things (IoT) \cite{Christidis:2016}, health care \cite{Lunardi:2019icitst}, smart vehicles \cite{Singh:2018b}, smart cities \cite{Kushch:2019} and others~\cite{Merlini:2019}.

However, a number of challenges need to be addressed before the widespread uptake of this technology. Zheng \textit{et al.} \cite{Zheng:2017} highlight several such challenges, including scalability and privacy leakage. Furthermore, two key challenges that are specifically relevant to IoT applications are high latency \cite{Yasaweerasinghelage:2017} and high energy consumption from some consensus algorithms \cite{Bach:2018}. They are relevant because of the high amount of generated data, the need for low latency for specific applications, and the hardware constraints on most IoT devices.

SpeedyChain~\cite{MichelinMOB:2018}~\cite{Lunardi:2018}~\cite{Lunardi:2019}, which uses the appendable-block blockchain model, was specifically proposed to address these issues by employing a different block structure and network architecture. The block structure allows the insertion of multiple transactions in the blockchain at the same time, and the network architecture uses the concept of gateways that allows higher performance devices to process transactions in the blockchain ~\cite{Lunardi:2018}. Moreover, it uses the practical byzantine fault tolerance (PBFT) consensus algorithm for achieving energy efficiency ~\cite{Bach:2018}. Michelin \textit{et al.} ~\cite{MichelinMOB:2018} demonstrated promising performance improvements by adopting SpeedyChain for storing and managing sensor data collected from smart vehicles in the context of smart city applications. 

Smart contracts can benefit from the appendable-blocks blockchain model used in SpeedyChain. In specific scenarios, the insertion of transactions in parallel can increase the performance of smart contracts. Moreover, the addition of smart contracts gives flexibility to applications that can work on top of SpeedyChain. For example, smart contracts can be used to help in IoT security in the following ways \cite{Christidis:2016}: (\textit{i}) providing an IoT update service; (\textit{ii}) allowing a marketplace between devices; (\textit{iii}) management and control of an IoT network; and (\textit{iv}) delegated processing and workload balancing.

Due to those benefits, in this work, we propose a model to provide the smart contracts capability on the SpeedyChain architecture, called context-based for appendable-block blockchain. The model works by isolating a group of smart contracts - from the same context - in a single block. Each group can receive transactions  independently - in parallel - to the other groups , thus increasing performance. To validate our model, we present an evaluation focused on delegated processing for route calculation based on GPS data for IoT devices. Although we apply the context-based smart contracts model to one specific blockchain technology, the ideas presented in this paper can be generalized and used in different architectures.

This work is divided as follows. Section \ref{sec:background} discusses the background concepts to understand our proposed model and its implementation. In Section \ref{sec:context} we establish the context-based smart contracts model. Section \ref{sec:implementation} discusses the changes to SpeedyChain to include our context-based model. Section \ref{sec:evaluation} define our experiment and discuss the results. Finally, final considerations and future work are presented in Section \ref{sec:final}.

\section{Background}
\label{sec:background}

This section presents the background required to understand the model of smart contracts for appendable-block blockchain. Therefore, we discuss different concepts used to build the model proposed in this paper. First, we present the functions used in the paper, followed by immutable-block blockchain structure, main appendable-block blockchain concepts, and the adopted smart contracts definition. 

\subsection{Mathematical functions}

The functions we use throughout these paper are:

\begin{itemize}
    
    \item We will use function $p$ to extract element  $e$  from a tuple using a lambda function as presented in equation \ref{example_lambda}:
    \begin{equation}
    p_{e}(tuple) = (\lambda (T_{1},...,T_{e},...T_{n}) \rightarrow T_{e})\label{example_lambda}
    \end{equation}

   % $$ p_{e}(tuple) = (\lambda (T_{1},...,T_{e},...T_{n}) \rightarrow T_{e})$$ %\quad where \quad 1 \leq e \leq n
   
    As an example for \eqref{example_lambda}, considering $ t = (1,6,3)$ the operations $p_{1}(t),  p_{2}(t)$ and $p_{3}(t)$ will result in $1, 6$ and $3$ respectively.
    
    % Hash
    \item We will use $H(x)$ as a hash function that can receive any sequence of bits $x$ as input and output another sequence of bits \cite{Lee:2007}. The specific $H(x)$ function used here will be abstracted. Proprieties of a good hash function, as collision-free, pseudo-randomness and unpredictability will just be assumed as true. 
   
   % Recuperar PK de uma assinatura digital
    \item The $PK$ function receives a digital signature as input and returns the public key from an asymmetric cryptography scheme. For this work, we consider a cryptography scheme in which you can recover a public key directly from a digital signature \cite{Johnson:2001}. 
\end{itemize}

\subsection{Immutable-Block Blockchain}
\label{subsection:traditionalBlockchain}

A blockchain is a distributed ledger that permanently stores all transactions that brought the system to the current state \cite{Nakamoto:2008}. Transactions are stored in blocks, once a block is added to the blockchain it is immutable, hence we call this type of blockchain an immutable-block blockchain. It is distributed because the system will work based on a Peer-to-Peer (P2P) network \cite{MALATRAS:2015}, in which each node in the blockchain network will maintain a local copy of the blockchain. 

The system state is changed by one node using the consensus algorithm to add new blocks. That causes all nodes to change their local copy of the system state. The algorithm works as a pre-agreement of how the system can progress, it helps all nodes to converge to the same system state despite some nodes malfunctioning or acting in a malicious way. There are multiple consensus algorithms, like Proof of Work (PoW) \cite{Nakamoto:2008}, Proof of Stake (PoS) \cite{Peercoin:2019} and Practical Byzantine Fault Tolerance (PBFT) %\cite{buchman:2016}
\cite{Castro:1999}. The consensus algorithm will be abstracted in this work to the function $performConsensus$ that returns  $true$ if a proposed block is authorized to be added to the blockchain or $false$ otherwise. More details about consensus are discussed in \cite{Lunardi:2019}.

%vantagens
The behavior of individual nodes is not presumed as correct or honest. The consensus algorithm guarantee that even if part of the nodes work maliciously or incorrectly, the data inserted in the blockchain can be trusted, and the system will work correctly. This feature is one of the major benefits of the blockchain. There are other benefits such as auditability \cite{Ali:2019}, since all transactions are stored as a ledger the current state can be audited at any time; resilience \cite{Ali:2019}, as a result of blockchain, is distributed in a P2P network, in the event of fault of any node, the blockchain continues working.

\subsection{Appendable-Block Blockchain}
\label{subsection:speedychain}

The immutable-block blockchain is nowadays the most used data architecture in blockchains. It is used in important data ledger technologies such as Ethereum\cite{EthereumDoc:2019}, Bitcoin \cite{Nakamoto:2008} and Hyperledger Fabric \cite{hyperledgerArxiv:2018}. However, there are other architectures proposed by industry, like the Tangle architecture proposed by IOTA \cite{M_Biradar_2018}, and by academia, as SpeedyChain \cite{Lunardi:2018}, whose data structure is relevant to this work. 

SpeedyChain is designed for the context of IoT, where devices usually have low computing power and limited storage capacity. This limits the capability to use a blockchain because of the necessity to store the blockchain in the nodes and the computing power required for some consensus algorithms such as PoW. Furthermore, high communication latency is a key factor in some IoT applications, which is another limiting factor in the use of blockchain in this case. To mitigate these problems, SpeedyChain proposes:(\textit{i}) to use gateways with more processing power and storage to work as blockchain full nodes, while other IoT devices have to connect to those gateways to access the blockchain. This removes the burden of maintaining a full node for limited IoT devices; and (\textit{ii}), a mutable blocks architecture, referenced as appendable-block blockchain, where blocks can be expanded with new transactions. This approach allows the blockchain to expand appending transaction in multiple blocks in parallel, while immutable-block blockchain can insert new transactions just by the introduction of a new block\cite{Lunardi:2018}\cite{MichelinMOB:2018}\cite{Lunardi:2019}.

Formally, a generalization of the appendable-block blockchain architecture has a set of $n$ nodes $N = \{N_{1}, ..., N_{n}\}$. Nodes are gateways and IoT devices that generate data. Each $\{N_{i}\}$ has a pair of public/secret keys $(NPK_{i}, NSK_{i})$ respectively from an asymmetric cryptography scheme, where the public keys are accessible to every participant of the blockchain\cite{Lunardi:2019}.

%Formally for a generic appendable-blocks blockchain, not related to Figure \ref{fig:speedyblock}, considers a set of $n$ gateways $G = \{G_{1}, ..., G_{n}\}$ and a set of $p$ devices $D = \{D_{1}, ..., D_{p}\}$, where each device is connected to one of the $G$ gateways and usually generates data. Gateways maintain the blockchain and additionally control access to it from devices. Each $\{G_{i}\}$ and $\{D_{i}\}$ have a pair of public-secret keys,  ($GPK_{i}$, $GSK_{i}$) and ($DPK_{i}$, $DSK_{i}$) respectively, where the public keys are accessible to every participant of the blockchain. The set of appliances is the set of all devices and gateways $A = D \cup G$, we will use the pair $(APK_{i}$, $ASK_{i})$ as a reference to the pair of public-secret keys of any element of $A$.

%A block in this model is a tuple $(BH, BD)$. $BH$, the block header, is a tuple $(parenteHash, NPK)$, the $parentHash$ is the hash value of the previous block $BH$ and works as a pointer to that block, the genesis block is the only block without a $parentHash$, the $NPK$ is the public key of the block owner, the genesis block is an exception and it does not have a public key, for this case its $NPK$ value is $\emptyset$. The $BD$ is the block data composed of committed transactions that we describe below. A blockchain is a nonempty set of blocks.

The data structure in an appendable-block blockchain is a non-empty set $BC$ of blocks. Each block is a tuple $(BH, BL)$. $BL$, named block ledger, is a set of transactions that can be incremented as necessary and linked to a $BH$. The $BH$, named block header, is another tuple composed of $(ParentHash, NPK_{i}, T)$ with meta-data about the block. Those fields are:
\begin{itemize}
    \item $ParentHash$ is the result of $H(BH)$ of the block inserted in the blockchain before this one. It works as a pointer to the previous block.
    \item $NPK_{i}$ is the public key of a member of the set $N$, only one block can have a specific $NPK_{i}$. To enforce that the $\nexists \, x \, | \, x, \, b \in BC \wedge p_{2}(p_{1}(x)) = p_{2}(p_{1}(b))$ post-condition must be respected. The node that has the $NSK_{i}$ to the $NPK_{i}$ of a block is said to be the block owner, and only that node can append new transactions to that block.
    \item $T$ is the first transaction inserted in a block and the only one to be part of the block header, furthermore, this transaction is the first transaction signed by a pair of public/secret keys, the public key is the $NPK_{i}$ value. 
\end{itemize}

The $BC$ set will form a hash-linked list of blocks $B$ connected by their $ParentHash$ in the $BH$. 
% Two post-conditions need to be obeyed to keep the $BC$ as a linked list:
% \begin{itemize}
%     \item $\nexistsx\, | \, x,\, y,\, q \in BC \wedge H(p_{1}(x)) = p_{1}(p_{1}(y)) \wedge H(p_{1}(x)) = p_{1}(p_{1}(q))$, which guarantees that each block can be pointed by at most one other block.
%     \item $\nexistsx\, |\, x,\, y \in BC \wedge \, p_{1}(p_{1}(x)) = \emptyset \wedge p_{1}(p_{1}(y)) = \emptyset$, which restraints to the existence of just one block that points to no other block with the $ParentHash$, this block is called the genesis block and is always the first block existing in the $BC$. 
% \end{itemize}
 
A transaction withholds data generated by the nodes, the data content depends on the application and context. In the appendable-block blockchain a transaction is represented as a tuple of $(Data, PT, Sig)$, where: $Data$ is specific to the node generating data through the creation of the transaction; $PT$ is a the hash of the previous transaction inserted in the block, it works as a hash-link connecting the transactions in the block. If it is the first transaction in the block, then it will refer to the hash of the $BH$; and $Sig$ is a digital signature from the node originating the $Data$ \cite{Lunardi:2019}.

 Before presenting how appendable-block blockchain adds new blocks and appends transactions, we present the auxiliary functions $newBlock$ as shown in Equation \ref{eq:newBlock}, which  summarizes the creation of a new block filling the necessary fields, and function $lB$ (Equation \ref{eq:lB}) , which returns a block that has no other block with the $ParentHash$ in the header pointing to it. In practice, it is the last block created in the blockchain.

     \begin{align}
          \begin{array}{ll}
             newBlock(T, BC) = \\
             ((H(p_{1}(lB(BC))),  PK(p_{3}(T)), T ), \{\}) 
         \end{array}
         \label{eq:newBlock}
     \end{align}
     \begin{align}
     \begin{array}{ll}
         lB(BC) =  x | x \in BC \wedge\\ 
         (\nexists y \in BC \wedge      p_{1}(p_{1}(y)) = H(p_{1}(x)) \wedge x \neq y )
         \label{eq:lB}
     \end{array}
    \end{align}

To add a new block to the blockchain, function $addB(BC, T)$ (Equation \ref{addBlock})  is used. It creates a new block for a node and appends the new block to the blockchain if there is no other block with the same $NPK$ as the transaction signature public key $PK(p_{3}(T))$. The predicate $uniqueBlock$ (Equation \ref{uniqueB}) guarantees this requirement. %The created block does not have any transaction in its $BL$.

\begin{align} 
addB(\!BC, T)\!=\!
\left  \{
\begin{array}{lll}
         BC\ \cup\ newBlock(T, BC),\\
        \text{if } uniqueBlock\\ \\
        BC\text{, otherwise}
\end{array}
\right.
\label{addBlock}
\end{align}

\begin{align}
 uniqueBlock = \nexists x.x\!\in\! BC\!\wedge p_{2}(p_{1}(x))\!=\!P\!K\!(p_{3}(T))
 \label{uniqueB}
\end{align}

New transactions are generated by nodes with new $Data$ to be inserted in the blockchain. This operation is performed only if the node's public key $(NPK_{i})$ is present in a block header $BH$. Function $appendT$ (Equation \ref{appendT}) specify the insertion of a new transaction $T$ in a block $B$ that has a public key equal to the public key used in the transaction signature.
%When $p_{2}(T)$ is equal to $PreTHash(B)$, the append function is balbalba
%\begin{equation}
%    appendT(BC, T) = (BC - B) + updateB(B, T)
%\end{equation}
%otherwise, the transaction T is not appended.
\begin{align}
 appendT(\!BC, T)\!=\!
 \left \{
 \begin{array}{lll}
     (\!BC - B) + updateB(\!B, T),\\ 
     \text{if } p_{2}(T) = PreTHash(B) \\ \\
     BC, \text{otherwise} 
 \end{array}
 \right.
 \label{appendT}
\end{align}

% $$
% where\ B = x\, |\, x \in BC \wedge p_{2}(p_{1}(x)) = PK(p_{3}(T))
% $$

$updateB$ (Equation \ref{updateB}) is an auxiliary function to generate an updated block where transaction $T$ is appended. Function $PreTHash$ (Equation \ref{PreTHash}) returns the hash of a previous transaction appended to the block or the block $B$ header hash. This hash will be used to check if a transaction is pointing to the $PT$ field of the last transaction inserted to the block.

\begin{equation}
    updateB(B, T) = (p_{1}(B), p_{2}(B) \cup T)
    \label{updateB}
\end{equation}
\begin{equation}
PreTHash(B) = 
\left \{
\begin{array}{ll}
    H(p_{1}(B)),\text{ if }|p_{2}(B)| = 0\\ 
    H(lastT(B)), \text{otherwise} 
\end{array}
\right.
\label{PreTHash}
\end{equation}

% $$
% \begin{array}{ll}
% lastT(B) = x\, | \, x \in p_{2}(B) \wedge   (\nexists y. y \in p_{2}(B) \wedge p_{2}(y) = H(x)) 
% \end{array}
% \end{equation}

Algorithm \ref{alg:appendable-blocks} shows how the main operation works on this model for the insertion of new transactions in a continuous way.
The $mempool$ consists of a set of transactions submitted to the blockchain by multiple nodes, but not yet appended to the blockchain, this $mempool$ is shared by all nodes.  Function $poll$, on the $mempool$, returns one transaction of the set. Before processing a new transaction, it checks if a block with the public key of the signer exists through function $exists$ (line 6); if not, a new block is processed by the consensus algorithm (line 7) and if approved, a new block is inserted and broadcast to the network (lines 9 and 10).
Otherwise, the proposed transaction is processed by the consensus algorithm (line 12) and if accepted, it is appended to the block owned by the transaction signer (more details about this algorithm is described in \cite{Lunardi:2019}).

\begin{algorithm}[h]
\caption{Main operation for Appendable-block Blockchain}
 \label{alg:appendable-blocks}
 \begin{algorithmic}[1]
\STATE Result: BC //Updated state
\STATE Input: mempool, BC  //Original state
\STATE {}
\WHILE{(True)}
\STATE T = poll(MemPool)
\IF {(!exists(PK($p_{3}$(T))))}
\STATE     ConsensusResponse = performConsensus(B)
\IF{(ConsensusResponse)}
\STATE        broadcast(B)
\STATE             BC = addB(BC, T)
\ENDIF
\ELSE
\STATE     ConsensusResponse = performConsensus(T)
\IF{(consensusResponse)}
\STATE         broadcast(T)
\STATE         BC = appendT(BC, T)
\ENDIF
\ENDIF
\ENDWHILE
 \end{algorithmic}
 \end{algorithm}

\subsection{Smart contracts}
\label{smartcontracts}

There are different models for smart contracts implementation in blockchains. One of these models is the one used by Ethereum \cite{EthereumWP:2018}. In Ethereum, smart contracts are stored in the blockchain as special transactions. Those transactions are bytecode that can be processed by the Ethereum Virtual Machine (EVM). Each node in the Ethereum network has an EVM. Calls, like a program call, for a smart contract, are appended to the blockchain as transactions. The transaction contains the bytecode, representing the program call, to be processed in the context of a specific smart contract. %This bytecode is inputted in the virtual machine with the current state of the smart contract, this will generate a change in the smart contract state, which will be saved by the node. Any new calls for that smart contract are processed using the new state. During the consensus algorithm, like any transaction, all the nodes need to check if the call done with the smart contract is valid. This is done using their EVM if the call is invalid the block with it is rejected \cite{Zorzo2018}.

Throughout this work, we will refer to a generic virtual machine as a function $VM$, which works similarly to the EVM \cite{EthereumDoc:2019}, \textit{i.e.}, $VM(S, Data) = S'$. $VM$ receives two inputs: $Data$, which is the bytecode with operations for the virtual machine; and, $S$, which is a pointer to a data structure that contains a state for the virtual machine and multiple smart contracts. The output for the $VM$ function is a reference to new state $S'$, based on the modifications that the $Data$ incurred. We store different states in a Merkle Patricia Trie \cite{EthereumWP:2018}. A new $S$ with no modifications is refereed as the constant $newS$.

\section{Context-based Model}
\label{sec:context}

This section presents the context-based model for extending the appendable-block blockchain architecture with the smart contracts capability. In this model, each block can hold a block state, which is a data structure capable of holding a mapping from an address to a smart contract state and code. Remember from the previous section that we use as a reference for the data structure a Merkle Patricia Trie\cite{EthereumWP:2018}. Aside from these new blocks with this capability to hold the block state, the blocks that carry just data from nodes, as presented in appendable-block blockchain (see Section \ref{subsection:speedychain}) still exist. During the creation of a new block it is decided which type of block will be created: a pure data block, which will carry just data; or, a block with context, which will hold a state.

\begin{figure}[h!]
    \centering
    \includegraphics[width=0.49\textwidth]{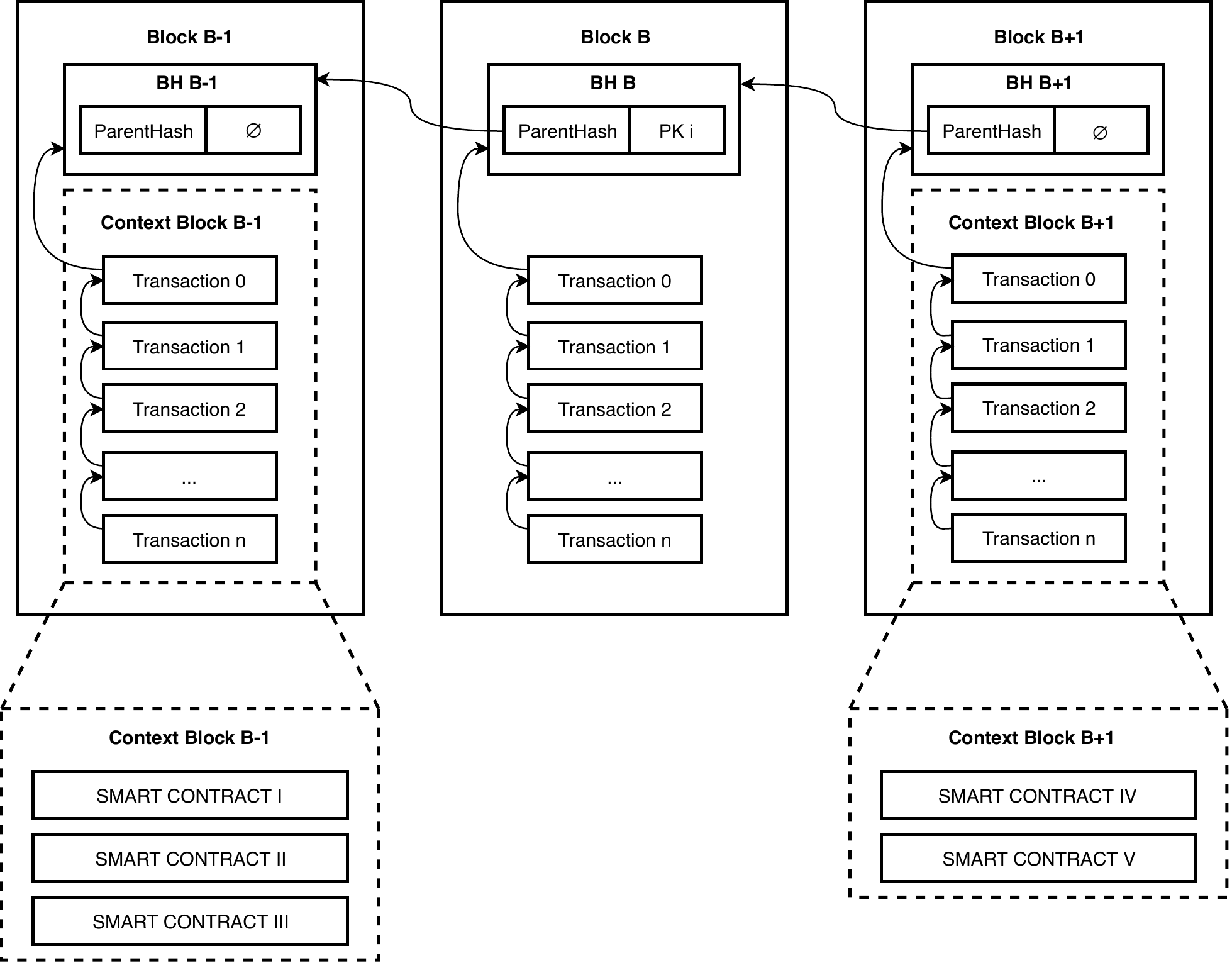}
    \caption{Smart contracts and Block context}
    \label{fig:smartcontracts}
\end{figure}

Figure \ref{fig:smartcontracts} presents an overview of the model. In the figure, three blocks are presented, only blocks $B-1$ and $B+1$ hold smart contracts, identified by the absence of a block $PK_{i}$. Therefore they are blocks with context. While block $B$ is a pure data block, identified by the presence of a $PK_{i}$. A context can have smart contracts, those smart contracts are isolated from other block context and can only interact with smart contracts in the same block, thus this model is called context-based model. In Figure \ref{fig:smartcontracts}, Smart Contract I can interact with Smart Contract II and III, this includes making a call to the other smart contracts, changing their states and querying information. Smart Contract IV, in block $B+1$, cannot interact with Smart Contract I, II and III in any form. A blockchain that adopts the proposed model can have any amount of blocks with context. The transactions stored in those blocks will carry a bytecode that represents a call for a smart contract or a command to create a new smart contract.

%A block state is the context that can be sent as input to the VM, it may contain smart contracts deployed in it and hold their states, a VM uses the $Data$ to operate on the existing smart contracts or deploy new ones. Those block states are exclusive to each block and isolated from the block state of the other blocks, therefore each block state is an isolated context for that block as presented in Fig. \ref{fig:smartcontracts}, in its block (B-1) and (B+1) have contexts on their own, while block (B) is a data block. If a user submits an operation to a smart contract in the context of the block (B-1) it can not interact with a smart contract in the context of the block (B+1) and neither access data appended to block (B).

All the functions in an appendable-block blockchain work exactly as previously presented in Section \ref{subsection:speedychain}, unless stated otherwise. The elements of the tuple $(BH,\, BL)$ representing the blockchain are different. $BH$ is a tuple $(ParentHash, Index, NPK, CTransaction)$ where $ParentHash$ and $NPK$ work as in the appendable-block architecture, although when the block has a context the value of $NPK$ will be equal to $\emptyset$, because there is no block owner or restriction of devices who can operate in a block. The  $Index$  field is a natural number whose value corresponds to the order of blocks created in the blockchain. The $CTransaction$ field stores a \textbf{committed transaction}, it is different from a \textbf{transaction}. A node can check if a block $B$ has a context by using function $HasContext$ (Equation \ref{HasContext}), and get a specific block by the index using function $getBlock$(Equation \ref{getBlock}).

\begin{equation}
HasContext(B) =
\left\{
    \begin{array}{ll}
        true  & \mbox{if } P_{3}(P_{1}(B)) = \emptyset \\
        false & \mbox{if } P_{3}(P_{1}(B)) \neq \emptyset
    \end{array}
\right.
\label{HasContext}
\end{equation}

\begin{align}
getBlock(Index) = x | x \in BC \wedge P_{2}(x) = Index
\label{getBlock}
\end{align}

A \textbf{committed transaction} is defined as a tuple $(Data, Sig, PT, BlockState)$, it is originated from a transaction that was processed by a node. The tuple fields are:
\begin{itemize}
    \item ${Data}$ is a binary sequence that depending on the block type will be treated differently. If it is a pure data block, then $Data$ represents data generated by a node, which will not be processed as $bytecode$. If it is a block with a block state, then it is a $bytecode$ that will be inserted as an input in the $VM$ function.
    \item $Sig$ is the digital signature of the transaction that generated this committed transaction.
    \item $PT$ is the hash value of the previous inserted committed transaction or the block header if it is the second committed transaction inserted the value will be the hash of $BH$.
    \item $BlockState$ is a pointer to a data structure holding the block state, as the Merkle Patricia Trie. The last committed transaction in the block has the most updated state, its value is generated by $VM$.
\end{itemize}

When a  $Data$ transaction includes bytecode to a block with context, it will be processed by function $VM$ (Equations \ref{VM}). %with as output will update the state to a new state and return a pointer to it. 
For that,  $Data$ and  $BlockState$ of the last inserted committed transaction in the block are inserted as input in the $VM$ function, the resulting pointer to a new state will be attached to a new committed transaction in the $BlockState$ field. $S'$ is a pointer to the resulting state, $S$ is the original BlockState and $NewS$ is an empty BlockState.

\begin{align}
VM(S, Data) =
\left\{
    \begin{array}{ll}
        S'  & \mbox{if } S \neq \emptyset  \\
        VM(NewS, Data) & \mbox{if } S = \emptyset
    \end{array}
\right.
\label{VM}
\end{align}

A node that wants to operate on the blockchain will create a transaction for that operation. The transaction is composed of $(Data, ToBlock, Sig, PT, OPCode)$, where the fields previously described in committed transactions are the same, the $ToBlock$ represents the destiny block where this transaction is to be processed. If the $ToBlock$ value is equal to $\emptyset$ and $OPCode$ is a specific value, then the transaction intention is to create a new block with a context or a pure data block. $OPCode$ is an integer that represents a code for the transaction intention, where 1 means the transaction is to create a new pure data block, 2 means the transaction is to create a new block with a context, and 3 means it is a transaction to be appended in a block.

Two functions will be used to summarize the block creation, when $OPCode$ is 1 or 2: Function $NewCBlock$ (Equation \ref{newCBlock}), which creates a new block with a context starting from a $newS$; and, function $NewPDBlock$ (Equation \ref{newPDBlock}), which creates a new pure data block.

\begin{align}
    \begin{array}{ll}
        newCBlock(T, BC) = \\
        ((H(p_{1}(lB(BC))),
        p_{2}(p_{1}(lB(BC)))+1,\,\emptyset, NCT_{C}),\{\})
    \end{array}
    \label{newCBlock}
\end{align}

\begin{equation}
NCT_{C} = (P_{1}(T), p_{3}(T), p_{4}(T), VM(\emptyset,P_{1}(T)))
\end{equation}

\begin{equation}
\begin{array}{ll}
    newPDBlock(T, BC) = ((H(p_{1}(lB(BC))),\\
    p_{2}(p_{1}(lB(BC)))+1,\, PK(p_{3}(T)), NCT_{PD}),\{\})
\end{array}
\label{newPDBlock}
\end{equation}

\begin{equation}
NCT_{PD} = (P_{1}(T), p_{3}(T), p_{4}(T), \emptyset)
\end{equation}

%The $newCBlock$ function create a new block with a context and add it to the blockchain, all new blocks include the first committed transaction which is relative to the transaction that created the block. The similar function $newPDBlock$ creates a pure data block. We define them as:
%The value $CT$ is the first committed transaction to a block with a context, that represents the transaction used for the block creation and can include a bytecode to be processed. It will introduce the first block state that will be used by the next committed transaction.

A transaction that has $OPCode$ 3 will be appended to a block. However, the transaction appended is treated differently if the intention is to append a transaction in a pure data block or a block with context. The  $CT_{C}$ function (Equation \ref{CTC}) creates a committed transaction to be appended in a block with a context. On the other hand, function $CT_{PD}$ (Equation \ref{CTPD}) will create a transaction to a pure data block. 

\begin{equation}
CT_{C}(T, B) = 
\left\{
    \begin{array}{lll}
        (P_{1}(T), p_{3}(T), p_{4}(T),\\ VM( p_{4}(lastCT(B)))),\\
        \text{if } p_{4}(T) = preCTH(B) \\
        \\
        B, \mbox{otherwise}
    \end{array}
\right.
\label{CTC}
\end{equation}

\begin{equation}
CT_{PD}(\!T, B)\!=\!
\left\{
    \begin{array}{lll}
        %\bullet 
        (P_{1}(T), p_{3}(T), p_{4}(T), \emptyset, P_{1}(T))) \\
        \text{if } p_{4}(T) = preCTH(B) \\
        \\
        %\bullet 
        B, \mbox{otherwise}
    \end{array}
\right.
\label{CTPD}
\end{equation}

\begin{equation}
preCTH(B) = 
\left \{
\begin{array}{ll}
    H(p_{1}(B)),\text{ if }|p_{2}(B)| = 0\\ 
    \\
    H(lastCT(B)), \text{otherwise} 
\end{array}
\right.
\label{preCTH}
\end{equation}

\begin{equation}
\begin{array}{ll}
lastCT(B) = \\x\, | \, x \in p_{2}(B) \wedge   (\nexists y| y \in p_{3}(B) \wedge p_{3}(y) = H(x)) 
\end{array}
\label{lastCT}
\end{equation}

The functions in equations \ref{CTC} and \ref{CTPD} are used in function $AppendT$ (Equation \ref{AppendT}), which directs a transaction to the correct function type by their $OPCode$ and updates $BC$.

\begin{equation}
AppendT(BC, T) =
\left\{
    \begin{array}{lllll}
        (BC - B) \cup CT_{C}(T, B)  \\ 
        \mbox{if } HasContext(B) = true  \\ \\
        (BC - B) \cup CT_{PD}(T, B)  \\  
        \mbox{if } p_{3}(B) = PK(p_{3}(T))  \\ \\
        BC, \mbox{ otherwise}
    \end{array}
\right.
\label{AppendT}
\end{equation}

\begin{equation}
B = getBlock(P_{2}(T))
\label{B}
\end{equation}

The algorithm for the main operation for this model is presented in Algorithm \ref{alg:contextalgorithm}. In the algorithm, $memPool$ works exactly like presented in Section \ref{sec:background}. Line 6 checks if the transaction being processed wants to append a new transaction to the blockchain ($OPCode$ 3) and if the destination block exists. If both are true, then the transaction is processed by the consensus algorithm and appended using the $appendT$ function (line 10). When the destination block does not exist, then a transaction creates a new block with a context (line 11). It checks the $OPCode$ value is equal to 2 and if the destination block is equal to $emptyset$, case both conditions are true then a new block will be created (line 15) after begin processed by the consensus algorithm. Finally (line 16), the algorithm checks if the transaction is to create a new pure data block ($OPCode$ 1). If so, then it is checked if there is no other block with the same public key as the signature, then it proceeds to create a new pure data block (line 20).

\begin{algorithm}[h]
\caption{Main operation for appendable-block blockchain with context-based model}
\label{alg:contextalgorithm}
\begin{algorithmic}[1]
%\begin{lstlisting}
\STATE {Result: BC //Updated state}
\STATE {Input: memPool, BC  //Original state}
\STATE {}
\WHILE {(True)}
\STATE {T = poll(memPool)}
\IF    {(exists($p_{2}$(T)) AND $p_{4}$(T) = 3)}
\STATE        {ConsensusResponse = performConsensus(T)}
\IF        {(consensusResponse)}
\STATE            broadcast(T)
\STATE            BC = appendT(BC, T)
\ENDIF
\ELSIF {($p_{2}$(T) = $\emptyset$ AND $p_{4}$(T) = 2 )   }
\STATE        ConsensusResponse = performConsensus(B)
\IF            {(ConsensusResponse)}
\STATE                {broadcast(B)}
\STATE                {BC = newCBlock(T, BC)}
\ENDIF
\ELSIF    {(!exists(PK($p_{3}$(T))) AND $p_{4}$(T) =1)   }
\STATE        ConsensusResponse = performConsensus(B)
\IF            {(ConsensusResponse)}
\STATE                broadcast(B)
\STATE                BC = newPDBlock(T, BC)
\ENDIF
\ENDIF
\ENDWHILE
%\end{lstlisting}
\end{algorithmic}
\end{algorithm}

\section{Implementation}
\label{sec:implementation}

\begin{figure}[h!]
    \centering
    \includegraphics[width=0.45\textwidth]{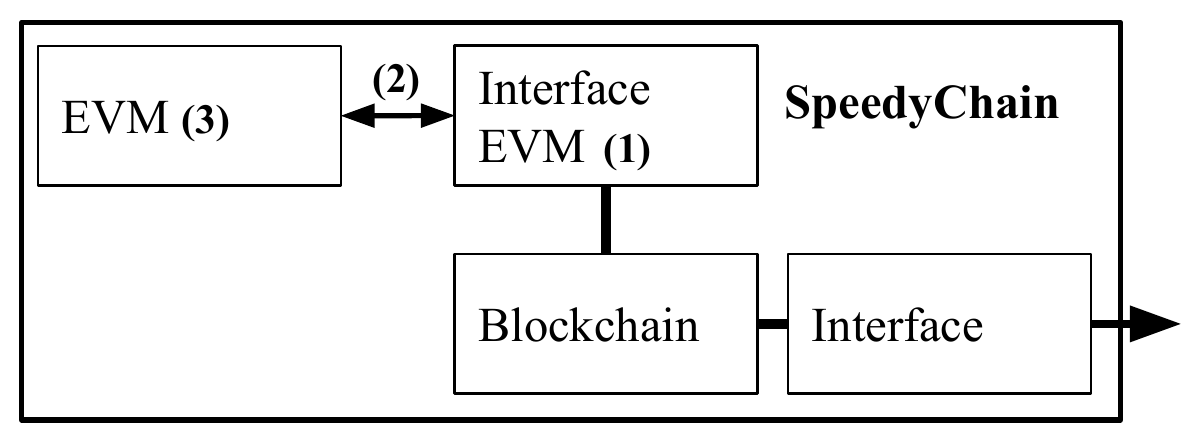}
    \caption{EVM and SpeedyChain}
    \label{fig:pimplmenetation}
\end{figure}

The implementation of the model should be easy to maintain and also easy to incorporate into the existing blockchain technology, in this paper into SpeedyChain. Taking that into consideration, we designed an ideal node architecture (see Figure \ref{fig:pimplmenetation}). Three new components are inserted in the SpeedyChain framework: Interface EVM (1), which is an interface in the SpeedyChain to communicate, through an inter-process communication protocol (2), with an internal EVM (3).  

% \begin{figure}[h!]
%     \includegraphics[width=0.5\textwidth]{figs/SAIR.pdf}
%     \caption{Current implementation}
%     \label{fig:rimplmenetation}
% \end{figure}

The process to run a smart contract starts with a new proposed transaction, which for this explanation contains a bytecode. We assume that all fields of the transaction and the blockchain are correct to receive this generic transaction. As presented in the model, Section \ref{sec:context}, this transaction targets an existing block. From this block, the last state is extracted and together with the bytecode is passed to the Interface EVM inside the SpeedyChain.

Interface EVM wraps both datas in a JavaScript Object Notation (JSON) format, which is then sent through an inter-process to the EVM. After sending the JSON, it awaits a return from the EVM with a result. The EVM receives the JSON with the state and bytecode. It changes the virtual machine state to the state received and using the bytecode as input process the request, which yields a resulting bytecode and an updated state. Both are wrapped in a new JSON object and sent as a response to Interface EVM. An error message is sent, if there is any problem when running the smart contract.

The results are unwrapped by the Interface EVM and handed to the blockchain, which will use the updated state to create a committed transaction as described in Section \ref{sec:context}. If an error is returned, then the transaction is discarded, and no alteration is applied to the blockchain.

% However, due to time constraints, it was not possible to implement the Ideal Design. As a result, the actual implementation uses the architecture in Figure \ref{fig:rimplmenetation}. There is the addition of the Device Simulator module, which assumes the function of the Interface EVM, which was removed. Additionally, it works generating data to be appended to the Blockchain, simulating to what a real device would generate. This Module resides outside of the SpeedyChain. All transactions generated by it are treated in SpeedyChain as pure data insertions destined to a pure data block, described in Subsection \ref{subsection:speedychain}.

% There is no block with a context in this implementation. That is an important consideration because it diverges from the model proposed. For using smart contracts, the bytecode and system state are treated as device data and stored in the Blockchain as such. As a result, only the device generating the information could operate on his smart contracts, because they reside on the device's block. 

%The EVM is connected through sockets to the SpeedyChain. In communication, it is still used JSON objects. 
This approach allows using the EVM developed by the Ethereum Foundation. It is implemented in Golang and could not be integrated internally with the SpeedyChain, implemented in Python. The use of that EVM implementation is important for maintainability. Any further update and modification by Ethereum Foundation in the EVM's operations would be inherited in our solution. %done to Geth's EVM by the Ethereum Foundation.
%As we use the Geth's EVM we will have prompt access to the updated EVM in our project.

%To compute a smart contract a process similar to the one from ideal design is used. Although, the Device simulator will operate as an intermediary and as a starting point. When the device simulator intents to process a smart contract it will check the last data inputted in its block and extract from it the last state. Then it will create a JSON with the bytecode it wants to be processed and the state extracted and send to the EVM, which will return a resulting bytecode and updated state. Lastly, the device simulator will commit a transaction to the Blockchain containing as data the returned bytecode and the state received.

\section{Evaluation} 
\label{sec:evaluation}

\begin{figure}[h!]
    \centering
    \includegraphics[width=0.4\textwidth]{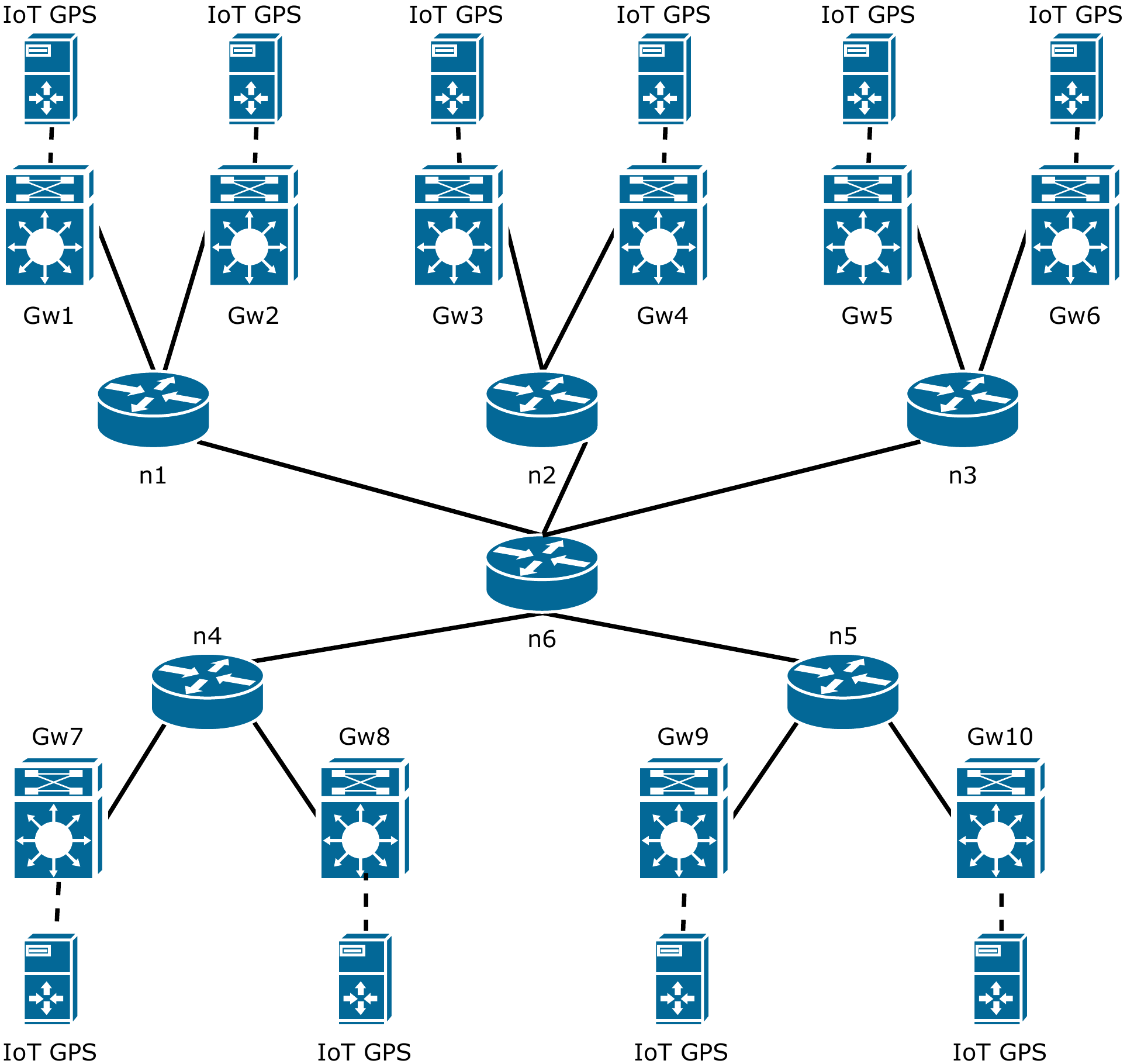}
    \caption{Emulated gateways architecture}
    \label{fig:core}
\end{figure}

In order to evaluate context-based smart contracts in appendable-block blockchains, we applied our solution in four testing  scenarios (see Table~\ref{tab:evaluation}). Every scenario was performed on a Virtual Machine (VM) with 4-core processor, 16GB of memory and 64MB of graphics memory running Ubuntu 18.04 operating system using a Virtual Box hypervisor over a Macbook Pro with 2.3 GHz 8-Core Intel Core i9 processor, 32GB DDR4 memory.%, and dual graphics boards (Intel UHD Graphics 630 and Radeon Pro 560X)
 In order to create a container-based network to emulate network equipment and gateways and devices the Core Emulator~\cite{CORE2008} was used. For all scenarios, a network with ten gateways was simulated (see Figure \ref{fig:core}). 

\begin{table}[h!]
\caption{Evaluated scenarios}
\label{tab:evaluation}
\begin{tabular}{ll}
\hline
\multicolumn{1}{c}{\textbf{Scenario}} & \multicolumn{1}{c}{\textbf{Description}}                                                                                                                                                                                                                                                          \\ \hline
A                                     & \begin{tabular}[c]{@{}l@{}}Sequential execution of 1,000 smart contracts without \\ external load\end{tabular}                                                                                                                                                                                    \\ \hline
B                                     & \begin{tabular}[c]{@{}l@{}}Sequential execution of 1,000 smart contracts with an \\ additional load of 50 devices per gateway and 100 \\ transactions per device, \textit{i.e.}, 5,000 transactions per \\ gateway and 50,000 transactions in total\end{tabular}                                           \\ \hline
C                                     & \begin{tabular}[c]{@{}l@{}}10 parallel context with 100 smart contracts transactions \\ per context (1,000 in total) without external load\end{tabular}                                                                                                                                           \\ \hline
D                                     & \begin{tabular}[c]{@{}l@{}}10 parallel context with 100 smart contracts transaction \\ per context (1,000 in total) with an additional load of \\ 50 devices per gateway and 100 transactions per device, \\ \textit{i.e.}, 5,000 transactions per gateway and 50,000 transactions \\ in total\end{tabular} \\ \hline
\end{tabular}
\end{table}

\textit{Scenario A} was performed to establish a baseline for smart contracts execution in appendable-block blockchains. One hundred smart contracts operations were executed to compute the distance between a device and a target using GPS information. We simulated ten devices with sequential computation of smart contracts operations (total of 1,000 sequential transactions) in the same gateway. Similarly to \textit{Scenario A}, \textit{Scenario B} performed the same number of sequential smart contracts computations, however, with an extra load of normal type of transactions in every gateway. A normal transaction in this experiment is a random input of data in a pure data block. The load was simulated through the insertion of 50 devices and 100 ``normal'' transactions produced by each device. In total, 50,000 additional ``normal'' transactions it were simulated. These transactions were produced by 500 devices.

Different from previous described scenarios, \textit{Scenario C} used the proposed context-based smart contracts in appendable-block blockchains. To do that, this scenario considered ten devices connected to ten different gateways and requesting 100 smart contracts operations (1,000 operations, the same computations as in \textit{A} and \textit{B}) in parallel, without any extra load in the gateways. Similarly to \textit{C}, \textit{Scenario D} adopted context-based smart contracts, with the same number of transactions operations as C, but with the same extra load present in \textit{B}.

\subsection{Metrics}\label{sec:metrics}

In order to perform an evaluation of the proposed context-based smart contracts in SpeedyChain, metrics called T1, T2, T3, T4, T5, and T6 were used:

\begin{itemize}
    \item \textbf{T1}: Time to perform consensus, insert a new block (first time that device is connected) and replicate it to all gateways; %T5 in the code
    \item \textbf{T2}: Time to add and replicate a device block to all gateways (after consensus); %T6
    \item  \textbf{T3}: Time to perform consensus, insert a new special block for smart contracts and update EVM with the new smart contract bytecode; %T11 in the code
    \item \textbf{T4}: Time to insert a transaction in the blockchain; %T1 in the code
    \item \textbf{T5}: Time to run a smart contract in the EVM and update the blockchain; %T14 in the code
    \item \textbf{T6}: Time to run all smart contracts evaluated (1,000 contracts operations).
\end{itemize}

All metrics are represented by an average time of ten repetitions for each scenario. A confidence level of 95\% was achieved.

\subsection{Results}\label{sec:results}

Context-based smart contracts execution can impact the consensus algorithms (\textbf{T1}) as presented in Table~\ref{tab:data-t1}. We can see that time to perform consensus can increase $\approx65$\% when using context-based smart contracts (\textit{scenario C}) when compared to a sequential smart contracts execution (\textit{scenario A}) in a scenario without an extra load in gateways. When comparing the scenario with extra load ("with normal transactions"), we can observe that consensus time is increased by less than 13\% when comparing with (\textit{Scenario D}) and without (\textit{Scenario B}) the context-based approach. This can be explained by the time required by the gateways to process smart contracts, affecting the time to perform consensus.

\small
\begin{table}[h!]
	\caption{Indicator \textbf{T1}: Time to perform consensus}
	\centering
    \begin{tabular}{c|c}
		\toprule
			\textbf{Scenario} & \textbf{Average time} \\
			\tiny{}            & \tiny{(in milliseconds)} \\
		\midrule
		    A          &  $19.54 \pm 0.24$	 \\
		    B   &  $90.98 \pm 0.77$ \\
            C  &  $32.40 \pm 1.43$ \\
            D          &  $102.67 \pm 0.87$ \\
	    \bottomrule
    \end{tabular}
	\label{tab:data-t1}
\end{table}

Also, we can observe similar behavior in the time to perform consensus and update all the gateways to ensure a global view of the blockchain (\textbf{T2}) presented in the Table~\ref{tab:data-t2}. In this case, we can observe a higher increment comparing scenarios \textit{C} to \textit{A} ($\approx85$\%), and \textit{D} to \textit{B} ($\approx14$\%). This result shows that it also affects not only the leader of the consensus but also all the gateways.

\small
\begin{table}[h!]
	\caption{Indicator \textbf{T2}: Time to perform consensus and replicate a device block to all gateways}
	\centering
    \begin{tabular}{c|c}
		\toprule
			\textbf{Scenario} & \textbf{Average time} \\
			\tiny{}            & \tiny{(in milliseconds)} \\
		\midrule
		    A          &  $46.53 \pm 0.48$	 \\
		    B   &  $279.64 \pm 2.53$ \\
            C  &  $85.59 \pm 3.55$ \\
            D          &  $320.24 \pm 2.93$ \\
	    \bottomrule
    \end{tabular}
	\label{tab:data-t2}
\end{table}

Table~\ref{tab:data-t3} presents the time to insert special blocks for a smart contract (\textbf{T3}). In all scenarios, the time required was higher than the average presented in \textbf{T2}. That can be explained by the communication and processing performed in the EVM. Additionally, it was observed a lower difference when comparing \textit{D} to \textit{B} ($\approx11$\%), than comparing \textit{C} to \textit{A}($\approx81$\%). 

\small
\begin{table}[h!]
	\caption{Indicator \textbf{T3}: Time to add block (in all gateways) and to update the EVM with the new smart contract bytecode)}
	\centering
    \begin{tabular}{c|c}
		\toprule
			\textbf{Scenario} & \textbf{Average time} \\
			\tiny{}            & \tiny{(in milliseconds)} \\
		\midrule
		    A          &  $61.24 \pm 0.65$	 \\
		    B   &  $305.22 \pm 33.45$ \\
            C  &  $111.03 \pm 4.45$ \\
            D          &  $340.03 \pm 40.04$ \\
	    \bottomrule
    \end{tabular}
	\label{tab:data-t3}
\end{table}

Time to insert transactions in the blockchain (\textbf{T4}) was less affected than block insertion by the proposed context-based smart contracts solution, as presented in the Table~\ref{tab:data-t4}. Scenario \textit{C} increased $\approx23$\% over \textit{A}, and \textit{D} increased $\approx7.5$\% over \textit{B}. Comparing the usage of context-based smart contracts adoption in \textbf{T3} and \textbf{T4}, we can observe that the impact in mean time to insert transactions (\textbf{T4}) was lower than in block insertion (\textbf{T3}).

\small
\begin{table}[h!]
	\caption{Indicator \textbf{T4}: Time to insert a new transaction in the blockchain}
	\centering
    \begin{tabular}{c|c}
		\toprule
			\textbf{Scenario} & \textbf{Average time} \\
			\tiny{}            & \tiny{(in milliseconds)} \\
		\midrule
		    A          &  $0.97 \pm 0.003$	 \\
		    B   &  $2.81 \pm 0.008$ \\
            C  &  $1.20 \pm 0.10$ \\
            D          &  $3.02 \pm 0.01$ \\
	    \bottomrule
    \end{tabular}
	\label{tab:data-t4}
\end{table}

One important measure is how much the processing of smart contracts is affected by individual executions (\textbf{T5}). We can observe in Table~\ref{tab:data-t5} that scenario \textit{C} increased in $\approx18$\% over \textit{A} and scenario \textit{D} increased in $\approx23$\% over \textit{B}. This shows that the parallel approach proposed in context-based smart contracts affect individual insertions. It can be explained by the larger number of messages exchanged by nodes. Although, 79.58ms (in the worst case) to receive the result of computation still very good considering the usual GPS update rate of one update per second (1Hz)~\cite{Huang:2012}.

\small
\begin{table}[h!]
	\caption{Indicator \textbf{T5}: Time to run a smart contract in the EVM, update the blockchain and return the result to device}
	\centering
    \begin{tabular}{c|c}
		\toprule
			\textbf{Scenario} & \textbf{Average time} \\
			\tiny{}            & \tiny{(in milliseconds)} \\
		\midrule
		    A          &  $29.17 \pm 0.03$	 \\
		    B   &  $64.18 \pm 0.40$ \\
            C  &  $34.47 \pm 0.10$ \\
            D          &  $79.58 \pm 0.72$ \\
	    \bottomrule
    \end{tabular}
	\label{tab:data-t5}
\end{table}

\begin{figure}[!h]
    \centering
    \includegraphics[width=0.49\textwidth]{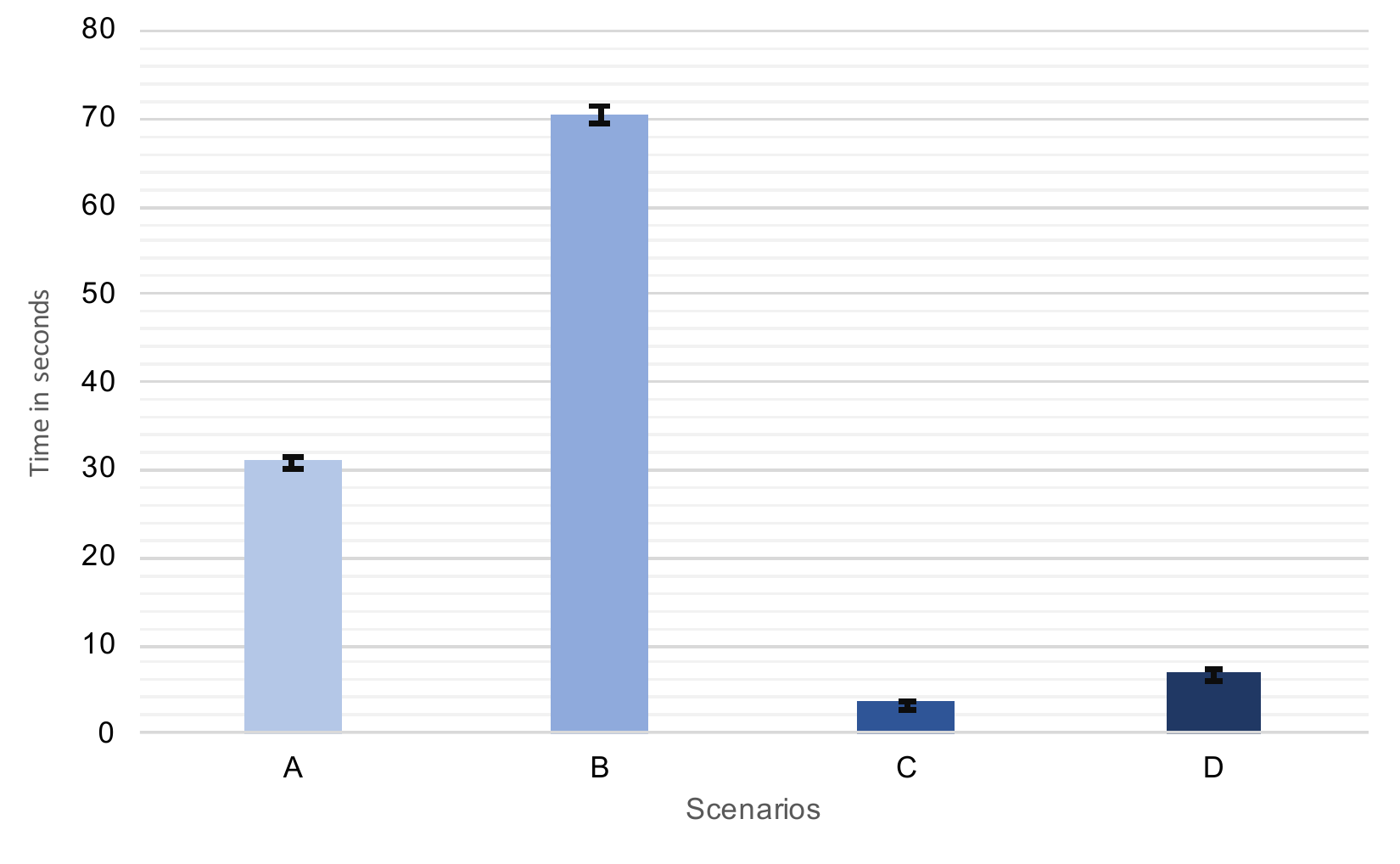}
    \caption{\textbf{T6}: Time to execute 100 calls for each 10 different smart contracts}
    \label{fig:t6}
\end{figure}

Finally, as presented in Figure~\ref{fig:t6}, time to perform calls for every smart contract (\textbf{T6}) was reduced when using context-based smart contracts in parallel execution (scenarios \textit{C} and \textit{D}) compared to ``traditional" sequential execution (scenarios \textit{A} and \textit{B}). Differently to the other metrics, \textbf{T6} is presented in seconds for each scenario. Scenarios \textit{A} and \textit{B} required an average of $31.22s{\pm0.43s}$ and $70.40s{\pm0.97s}$, respectively. Although, scenarios \textit{C} and \textit{D} required only $3.46s{\pm0.12s}$ and $6.96s{\pm0.48s}$. While context-based approach (\textit{C} and \textit{D}) presented an impact in the main operations (\textbf{T1}, \textbf{T2}, \textbf{T3}, \textbf{T4} and \textbf{T5}) of the blockchain, in \textbf{T6} required around 10\% of the time to perform all smart contracts  than in the sequential approach (scenarios \textit{A} and \textit{B}).

% \small
% \begin{table}[h!]
% 	\caption{Indicator \textbf{T6}: Time to execute all calls to smart contracts (100 calls for each of 10 smart contracts)}
% 	\centering
%     \begin{tabular}{c|c}
% 		\toprule
% 			\textbf{Scenario} & \textbf{Average time} \\
% 			\tiny{}            & \tiny{(in seconds)} \\
% 		\midrule
% 		    A          &  31.22 \pm 0.43	 \\
% 		    B   &  70.40 \pm 0.97 \\
%             C  &  3.46 \pm 0.12 \\
%             D          &  6.96 \pm 0.48 \\
% 	    \bottomrule
%     \end{tabular}
% 	\label{tab:data-t6}
% \end{table}

\subsection{Discussion and threats to validity}\label{sec:discussion}

Results presented in the Evaluation section show the possible gain when using context-based smart contracts. Reducing the total time of smart contracts can contribute to a more distributed execution of smart contracts than the current solutions. The adoption of appendable-block blockchain allowed a parallel execution, resulting in smaller time to perform smart contracts operations (as presented in \textbf{T6}). 

Nonetheless, overhead is introduced by context-based smart contracts, especially when comparing a scenario without an extra load on gateways. Context-based approach increased in more than 60\% the time to achieve consensus for new blocks in the blockchain (\textbf{T1}, \textbf{T2} and \textbf{T3}) in scenario without extra load. However, scenarios with extra load (simulating a real blockchain scenario) presented a small increase for the same indicators (less than 15\%).  Also, it is important to note that the block insertion is performed only once for each context.

There are some threats to the validity on the results presented in Section \ref{sec:results}. The first threat is related to hardware capability. In this work, we did not used physical IoT devices and GPS. However, we used the same cryptography algorithms and methods than those that were adopted by Lunardi \textit{et al.}~\cite{Lunardi:2018} in their experiments (using real hardware). Consequently, devices using IoT hardware should be capable of executing the same operations, but probably with different performance.

Additionally, the evaluation did not consider possible signal and communications problems. Consequently, the experiments did not take into account issues that mobile devices and/or gateway can produce. Hence, this threat should be further discussed and mitigated in a future work.

Finally, threats related to security issues that the proposed approach can introduce. For example, the impact of tampered devices producing invalid data (\textit{e.g.}, invalid coordinates) were not considered. This specific threat should also be better addressed in future work. 

\section{Final Considerations \& Future Work}
\label{sec:final}

In this paper, we presented a new model for context-based smart contracts that can be applied to appendable-block blockchains. This model expands the possibilities that the appendable-block blockchains can be used for. Also, it can allow improvements in the performance of smart contracts computation through parallel execution.

Results presented in the evaluation indicate that the execution of smart contracts can be reduced when they are classified in independent contexts. For example, in the scenario with additional load, the time to perform all the smart contracts computations was reduced from more than 70 seconds to less than 7 seconds. However, the proposed context-based smart contracts increased the time to perform consensus, due to the usage of gateways' processors to compute smart contracts, although, block insertion is performed only once per context. Additionally, in the scenario with extra load, this increase was reduced when compared to a scenario without extra load.

Due to space limitation, in this paper we did not discuss security issues that could be introduced by this new model.  For  example,  there is  no  protection  against  replay  attacks, an  attack  where  a transaction  is  copied  by  a  malicious  user  and  sent  to be  processed  again.

%;  Second  it  is  missing  a  performance  evaluation,  although  it  would  be  more complicated than  in  other  blockchains  because  of  the  parallel  nature  of  analysis  based  on  just how  much the transactions  can  be  processed  would  not  suffice, the results would greatly change based on the application and its characteristics.

As future work, we intend to evaluate the proposed approach using actual hardware, in special to evaluate the impact that both latency and constrained hardware can have in context-based smart contracts. Also, we expect to reduce the overhead introduced by the context-based approach. Additionally, we intend to evaluate possible security issues and possible new attacks that can be explored in this approach. Finally, we intend to expand the context-based smart contracts model to other blockchains, such as Ethereum and Hyperledger Fabric.

% \section*{Acknowledgment}




\bibliographystyle{IEEEtran}
\bibliography{bare_conf}


% \end{thebibliography}

\end{document}